\documentclass[]{spie}  

\usepackage{amsmath,amsfonts,amssymb}
\usepackage{graphicx}
\usepackage[colorlinks=true, allcolors=blue]{hyperref}
\usepackage{xcolor}
\usepackage[export]{adjustbox}

\title{MAVIS: Enabling High-Precision Ground-Based Astrometry in the Visible Spectrum}

\author[a,i]{Mojtaba Taheri}
\author[b,c]{Jesse Cranney}
\author[d]{Antonino Marasco}
\author[e]{Stephanie Monty}
\author[f]{Davide Massari}
\author[g]{Guido Agapito}
\author[g]{Giovanni Cresci}
\author[h]{Richard M. McDermid}
\author[b,c]{Francois Rigaut}
\author[i]{Benoit Neichel}
\author[b,c]{David Brodrick}
\author[g]{Cédric Plantet}

\affil[a]{Thirty Meter Telescope International Observatory, 100 West Walnut Street, Pasadena, California, United States}
\affil[b]{Advanced Instrumentation Technology Centre, Research School of Astronomy and Astrophysics, Australian National University, Canberra, Australia}
\affil[c]{Astralis Instrumentation Consortium, Australia}
\affil[d]{INAF - Padova Astronomical Observatory, Vicolo dell’Osservatorio 5, 35122 Padova, Italy}
\affil[e]{Institute of Astronomy, University of Cambridge, Madingley Rd, Cambridge, CB3 0HA, UK}
\affil[f]{INAF - Osservatorio di Astrofisica e Scienza dello Spazio di Bologna, Via Piero Gobetti 93/3, 40129 Bologna, Italy}
\affil[g]{INAF - Osservatorio Astrofisico di Arcetri, Largo E. Fermi 5, 50127 Firenze, Italy}
\affil[h]{Astrophysics and Space Technologies Research Centre, Macquarie University, Sydney, NSW 2109, Australia}
\affil[i]{Aix Marseille Univ, CNRS, CNES, LAM, 38 Rue Frédéric Joliot Curie, 13013, Marseille, France}

\definecolor{myred}{RGB}{255, 0, 0}
\begin{document} 
\maketitle

\begin{abstract}
MAVIS (the MCAO-Assisted Visible Imager and Spectrograph), planned for the VLT Adaptive Optics Facility, represents an innovative step in Multi-Conjugate Adaptive Optics (MCAO) systems, particularly in its operation at visible wavelengths and anticipated contributions to the field of astronomical astrometry. Recognizing the crucial role of high-precision astrometry in realizing science goals such as studying the dynamics of dense starfields, this study focuses on the challenges of advancing astrometry with MAVIS to its limits, as well as paving the way for further enhancement by incorporating telemetry data as part of the astrometric analysis. We employ MAVISIM, Superstar, and DAOPHOT to simulate both MAVIS imaging performance and provide a pathway to incorporate telemetry data for precise astrometry with MAVIS. Photometry analyses are conducted using the Superstar and DAOPHOT platforms, integrated into a specifically designed pipeline for astrometric analysis in MCAO settings. Combining these platforms, our research aims to elucidate the impact of utilizing telemetry data on improving astrometric precision, potentially establishing new methods for ground-based AO-assisted astrometric analysis. This endeavor not only sheds light on the capabilities of MAVIS but also paves the way for advancing astrometry in the era of next-generation MCAO-enabled giant telescopes.
\end{abstract}


\section{Introduction}

The MCAO-Assisted Visible Imager and Spectrograph (MAVIS), planned for installation at the European Southern Observatory (ESO) Very Large Telescope's (VLT) Adaptive Optics Facility (AOF), represents a significant advancement in adaptive optics (AO) instrumentation, specifically designed to operate at visible wavelengths. MAVIS is poised to be the first Multi-Conjugate Adaptive Optics (MCAO) instrument operating in the visible, offering unprecedented resolution and sky coverage for an 8m-class telescope \cite{mavis2020science,mavis2020SPIE}. The instrument's capabilities are expected to have a profound impact on astronomical observations, particularly in the realm of high-precision astrometry. This field is crucial for a wide array of astrophysical investigations, including studying the dynamics of dense starfields (e.g., towards the Galactic bulge, and globular cluster cores) \cite{mavis2020science,2021MNRAS.507.2192M}, characterizing exoplanetary systems through astrometric signatures \cite{ellis2020mavis}, and probing the structural properties and evolution of extra-galactic star clusters and their hosts \cite{mavis2021Msngr}.

Astrometry, the precise measurement of stellar positions and motions, relies heavily on advanced photometric analysis tools. DAOPHOT, a widely-used software for stellar photometry, has been a cornerstone in this field \cite{stetson1987daophot}. However, the increasing complexity of point spread functions (PSFs) in the MCAO era, coupled with their variations across images, presents challenges for traditional photometry tools \cite{massari2016high}. These complex PSFs are difficult to describe analytically, necessitating innovative solutions that incorporate additional information such as AO telemetry data to improve accuracy.

The SuperStar platform addresses these challenges by making it possible to take a PSF table as an input. This PSF table is made using the telemetry data recorded during AO observations and processed with PSF reconstruction platforms such as TIPTOP and PASSATA \cite{marasco2020superstar, agapito2016passata, neichel2021tiptop}. In addition to its capability to extract purely numerical PSFs from different regions of the image, SuperStar can use these PSFs to determine the positions and magnitudes of stellar sources. The iterative PSF fitting process refines the PSF models, astrometry, and photometry, adapting to the varying PSFs and delivering accurate measurements even in the presence of complex AO-corrected images. This approach provides a significant advantage over traditional photometric tools, which cannot incorporate telemetry data in this manner.

In this study, we utilize the MAVIS Image Simulator (MAVISIM), a software tool designed to simulate the imaging capabilities of the MAVIS instrument under various observing conditions, to generate a series of synthetic datasets that mimic the observational conditions of MAVIS \cite{monty2020mavis}. MAVISIM incorporates a comprehensive model of the MAVIS instrument, including its optical design, throughput, adaptive optics performance, and detector characteristics. It accounts for various sources of astrometric error, such as point spread function (PSF) variability due to high-order aberrations, tip-tilt residuals, and static distortions introduced by the AO module's opto-mechanics. To comprehensively evaluate the astrometric performance of MAVIS, we have designed multiple datasets with varying levels of atmospheric turbulence and different field densities (100 and 1000 stars). These datasets are then used to compare the astrometric performance of the DAOPHOT and Superstar platforms. The result of this study paves the way for the next step, which is leveraging Superstar's abilities to improve MAVIS capabilities by utilizing telemetry information.
\section{Methodology}

In this section, we outline the methodology used to evaluate the astrometric performance of MAVIS. Our approach involves using simulations to create various datasets, followed by a detailed analysis of the astrometric residuals compared to the ground truth for each dataset. We used MAVISIM to simulate six different scenarios to compare the astrometric recovery of DAOPHOT and Superstar in the context of MAVIS. These scenarios are listed in Table \ref{tab:scenarios}.

Scenarios S1 and D1 analyze a grid of 21x21 stars over the MAVIS field of view without considering atmospheric turbulence. This comparison investigates whether the measurement error floor for the combination of MAVISIM and both platforms is reasonably smaller than the measured error in realistic cases. The other four scenarios account for the comparison of astrometric performance in the presence of atmospheric turbulence and other sources of error for different field densities.

\begin{table}[h]
\centering
\caption{Simulation Parameters}
\label{tab:scenarios}
\begin{tabular}{c|c c c}

Scenario & {Number of Stars} & Photometric Analysis Platform & Atmospheric Turbulence \\ \hline
S1 & 441 & Superstar & No \\
D1 & 441 & DAOPHOT & No \\ 
S2 & 100 & Superstar & Yes \\ 
D2 & 100 & DAOPHOT & Yes \\ 
S3 & 1000 & Superstar & Yes \\ 
D3 & 1000 & DAOPHOT & Yes \\ 
\end{tabular}
\end{table}

\subsection{Simulation}

To simulate the observational conditions of MAVIS, we utilize the MAVIS Image Simulator (MAVISIM). This simulation tool is crucial for understanding how different factors affect the performance of MAVIS, particularly in terms of achieving high-precision astrometry. The main challenges to astrometric recovery in MAVIS are both simulated in MAVISIM, namely:
\begin{itemize}
\item \textbf{PSF Field Variability:} PSF variability is a significant source of astrometric error, especially in AO systems where the PSF can change across the field of view due to uncorrected high-order aberrations. MAVISIM ingests a database of PSFs which have been sampled across the MAVIS field of view. These PSFs are themselves generated by an end-to-end AO simulator, in our case, COMPASS\cite{ferreira2018compass}.
\item \textbf{Static Distortion:} MAVISIM also accounts for static distortions due to the opto-mechanics of the AO module. These distortions are mapped using end-to-end simulations in Zemax, providing a detailed static distortion map for the MAVIS AO system.
\end{itemize}

\begin{figure}[h]
\centering
\includegraphics[width=0.3\textwidth]{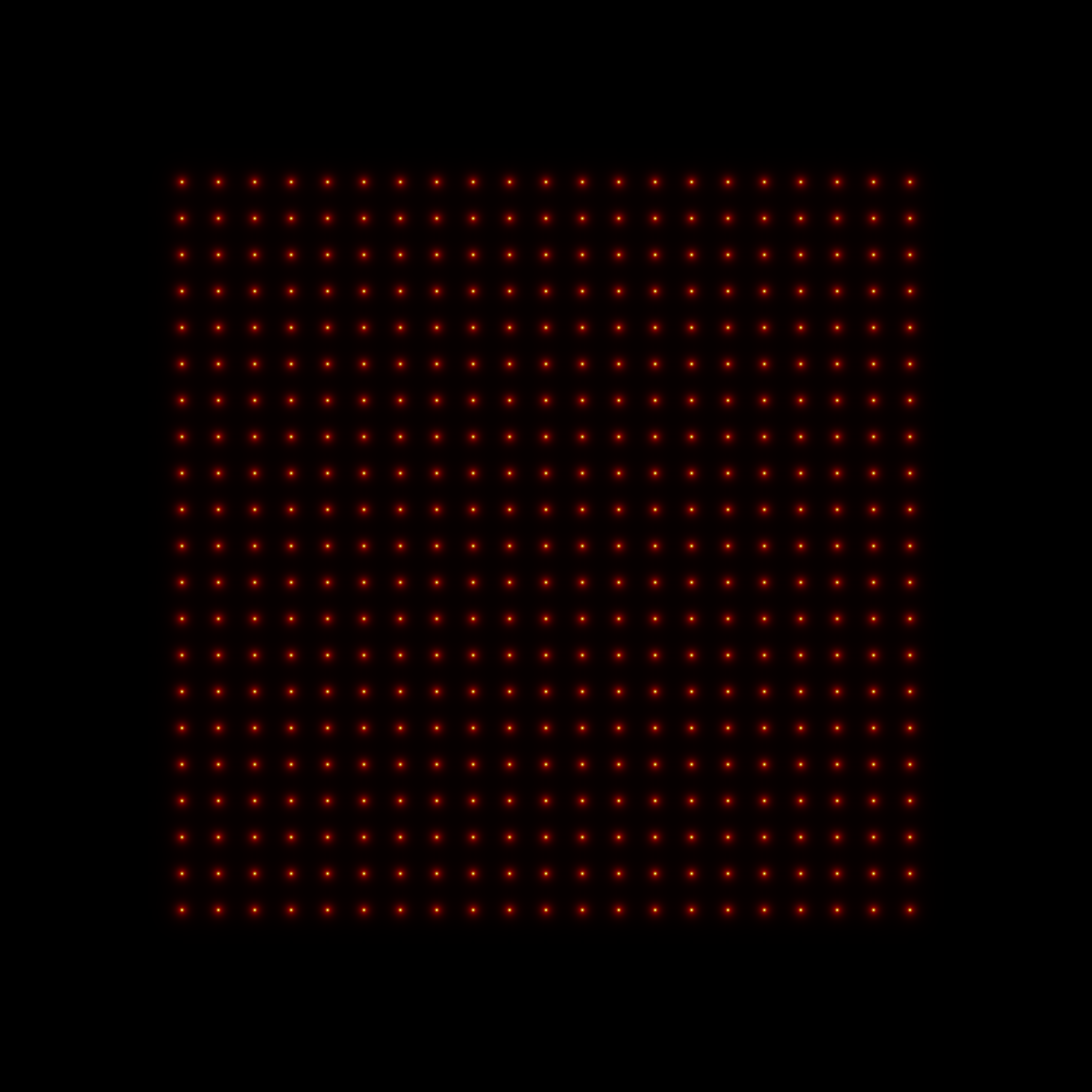}
\includegraphics[width=0.3\textwidth]{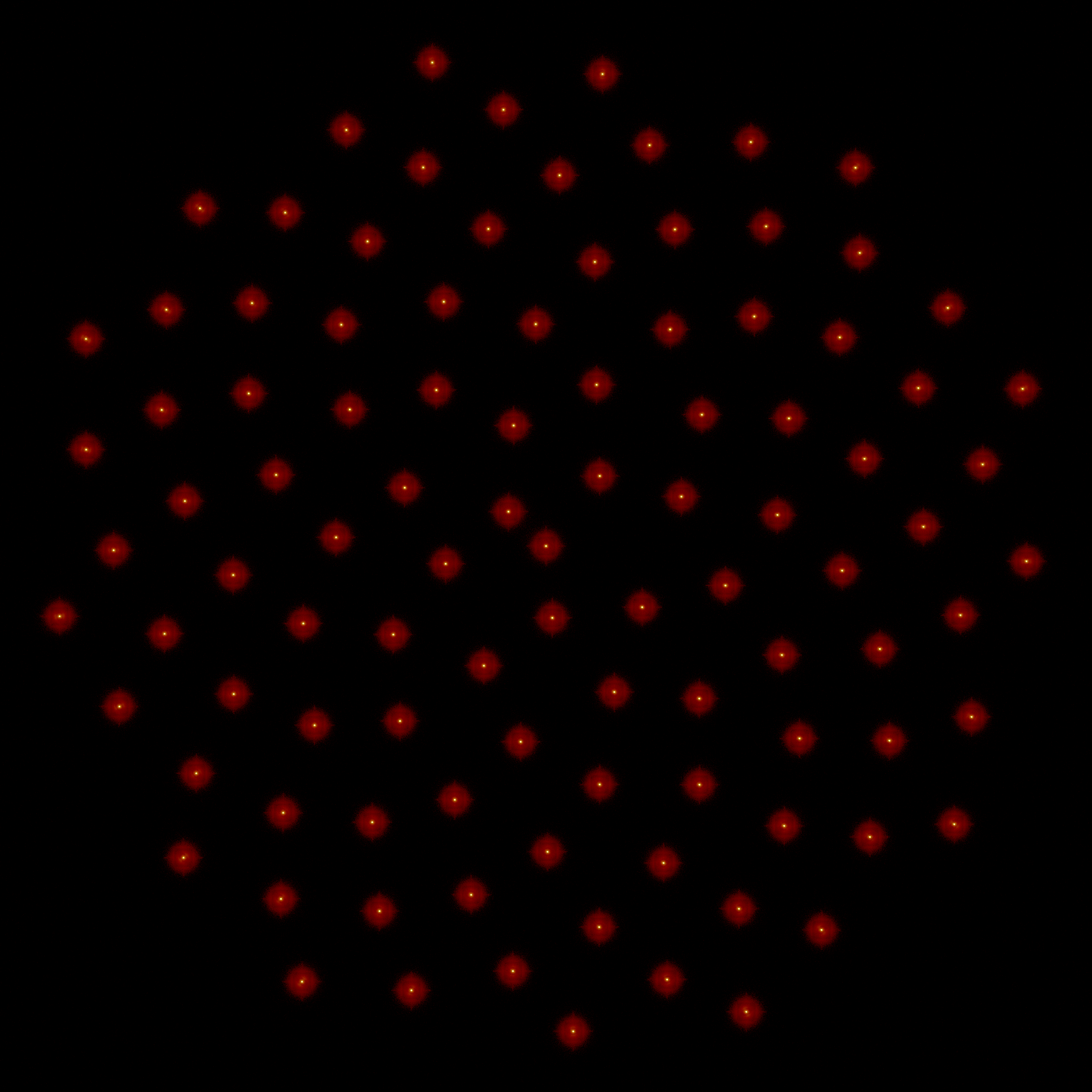}
\includegraphics[width=0.3\textwidth]{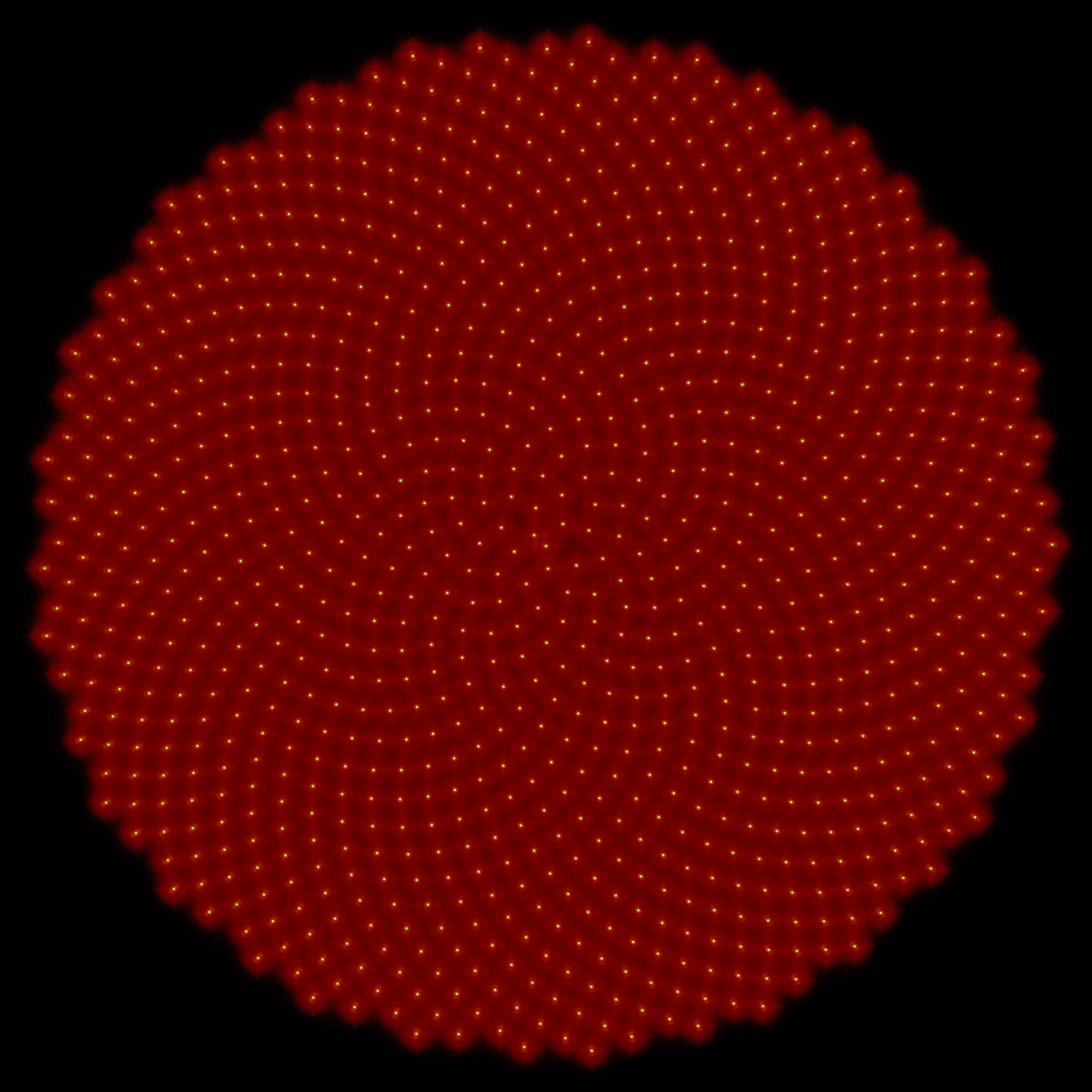}
\includegraphics[width=0.3\textwidth,trim={50cm 50cm 50cm 50cm},clip]{noiseless_grid.png}
\includegraphics[width=0.3\textwidth,trim={50cm 50cm 50cm 50cm},clip]{e2e_100_stars.png}
\includegraphics[width=0.3\textwidth,trim={50cm 50cm 50cm 50cm},clip]{e2e_1000_stars.png}
\caption{Simulated fields:  \{sanity check, 100 stars, 1000 stars\}. Top row: full field; bottom row: zoom on central star.}
\label{fig:simulated_field_100}
\end{figure}

Common simulation parameters are summarized in Table \ref{tab:simulation_parameters}. A representation of the simulated field for different scenarios is shown in Figure \ref{fig:simulated_field_100}.

\begin{table}[h]
\centering
\caption{Simulation Parameters}
\label{tab:simulation_parameters}
\begin{tabular}{lr}
Parameter & Value \\
\hline
\hline
Target wavelength & 550 nm\\
Frame-rate& 1000 Hz\\
Num. of atmospheric layers& 10 \\
Altitude (min, max) & $(30, 14000)$ m\\
Wind speed (min, max) & $(4.5, 34.3)$ m/s\\
Wind dir. (min, max)& $(0^{\circ}, 25^{\circ})$ \\
\hline
Num. of LGS& 8 \\
Constellation diam & 35''\\
Num. of sub-aperture& $40\times40$\\
Read-out noise & 0.2 $e^-$ \\
Flux/sub-ap/ms & 75 photons \\
\hline
Num. of NGS& 3\\
Constellation diam & 40''\\
Num. of sub-aperture& $1\times1$\\
Read-out noise & 0.5 $e^-$ \\
Flux/sub-ap/ms & 1200 photons \\
\hline
Num. of DMs & 3\\
Altitude & [0, 6, 13.5] km\\
Actuator pitch & [22, 25, 32] cm\\
\end{tabular}
\end{table}

Figures \ref{fig:simulated_field_100} show examples of the simulated fields for 100 and 1000 stars.

\subsection{Photometric Analysis}

We conducted astrometric analyses using both DAOPHOT \cite{stetson1987daophot} and Superstar \cite{marasco2020superstar}\footnote{a preliminary version of Superstar has been used in Ref. \citeonline{2020SPIE11448E..0GM}} for scenarios with and without atmospheric turbulence, across simulated stellar fields containing 100 and 1000 stars as listed in Table \ref{tab:scenarios}. The analysis involved extracting the PSF model in different regions of the field, using these PSFs to determine the precise positions of sources, and iterating this process to refine the PSF models and improve both astrometric and photometric accuracy. Parameters of the photometric analysis are listed in Table \ref{tab:analysis_parameters}.

\begin{table}[h!]
\centering
\caption{Parameters of the Photometric Analysis}
\label{tab:analysis_parameters}
\begin{tabular}{p{4cm}  p{2.7cm}  p{3.5cm}  p{5cm}}
Parameter & DAOPHOT & SuperStar & Notes \\ \hline
PSF Radius & 10~pixels & 30 pixels & Radius of circle within which PSF is defined \\
Fitting Radius & 4.10~pixels & 5 pixels & Radius within which the PSF is fit to every star \\
Order of PSF Variability & Quadratic & none (numerical PSF) &  \\ 
Analytical Form of PSF & Moffat ($\beta=1.5$) & none (numerical PSF) &  \\
Number of subregions & none & 3$\times$3 & The PSF is built separately in each subregion \\
($N_{\rm min}$,$N_{\rm max}$) & none & (3,15) & min and max number of stacked stars in each subregion\\
\end{tabular}
\end{table}

\subsection{Astrometric Analysis}

The astrometric analysis was performed using a custom pipeline, the details of which are published in \cite{taheri2022optimal}. This pipeline automates the process of matching the catalog of detected sources from the photometry analysis to the ground truth catalog, then calculates the best-fit affine transformation between the two catalogs, effectively accounting for any translation, rotation, and scaling differences.

To compensate for distortions in the field, the pipeline employs a fifth-degree polynomial model. This model absorbs the majority of the low-order distortions across the fields of view while leaving high-order information such as proper motion of stars or, in this case, position measurement uncertainties, untouched \cite{massari2016high}. By subtracting this distortion model from the best affine match, we obtain the astrometric residuals, which represent the remaining discrepancies between the measured and true positions of the stars. Ideally, these astrometric residuals should be zero in the absence of proper motion between the ground truth and the photometry catalogs. However, phase aberrations induced by atmospheric turbulence introduce errors in the determination of stellar positions, leading to non-zero residuals.

\section{Results}

In this section, we present the results of the astrometric performance analysis for MAVIS. We focus on the astrometric residuals and the root mean square (RMS) residuals for each scenario, comparing the performance of DAOPHOT and SuperStar under different conditions. The results are summarized in Table \ref{tab:results_scenarios}.

\begin{table}[h]
\centering
\caption{Astrometric Residuals for Different Scenarios}
\label{tab:results_scenarios}
\begin{tabular}{c c c}

Scenario & Platform & RMS astrometric Residual (millipixels) \\ \hline
S1  & SuperStar  &  1\\ 
D1  & DAOPHOT    &  0.4\\ 
S2  & SuperStar  &  24\\ 
D2  & DAOPHOT    &  23\\ 
S3  & SuperStar  &  20\\ 
D3  & DAOPHOT    &  17\\ 
\end{tabular}
\end{table}

The results demonstrate that both Superstar and DAOPHOT achieve comparable astrometric performance across different scenarios. The RMS residuals provide a measure of the overall precision of the astrometric measurements, indicating that Superstar performs similarly to DAOPHOT even without leveraging PSF reconstruction and telemetry data.

\section{Discussion and Conclusion}

The results from our study demonstrate that the astrometric performance of MAVIS analysed by Superstar is comparable to that of DAOPHOT. Table \ref{tab:results_scenarios} indicates that both platforms perform similarly across various scenarios simulated by MAVISIM. The comparable RMS residuals observed for Superstar and DAOPHOT provide foundation for future enhancements. Specifically, leveraging Superstar's ability to utilize a PSF table, reconstructed from telemetry data using platforms like TIPTOP and PASSATA, offers a promising path forward. This capability can potentially improve astrometric precision beyond the current results.

The successful demonstration of Superstar's performance underlines the importance of continued development and refinement of this tool. Integrating PSF reconstruction techniques and telemetry data into Superstar could lead to significant advancements in astrometric accuracy for MAVIS.

In summary, our study has shown that Superstar performs comparably to DAOPHOT without additional telemetry data. This opens up the possibility of further enhancing astrometric precision by incorporating PSF reconstruction from telemetry data in future work. The findings from this study highlight the potential of Superstar and pave the way for advancing ground-based astrometry with MAVIS.

\appendix

\section{Astrometric Reduction Plots}

In this appendix, we present the astrometric reduction plots for the six scenarios listed in Table \ref{tab:scenarios}. Each figure contains three panels: the left panel shows the distortion after the best affine match, the middle panel shows the degree five distortion model, and the right panel shows the astrometric residual. As expected, the remaining distortion exhibits high spatial frequency behavior, representing the remaining noise from the photometric analysis.

\begin{figure}[h!]
\centering
\includegraphics[width=1\textwidth]{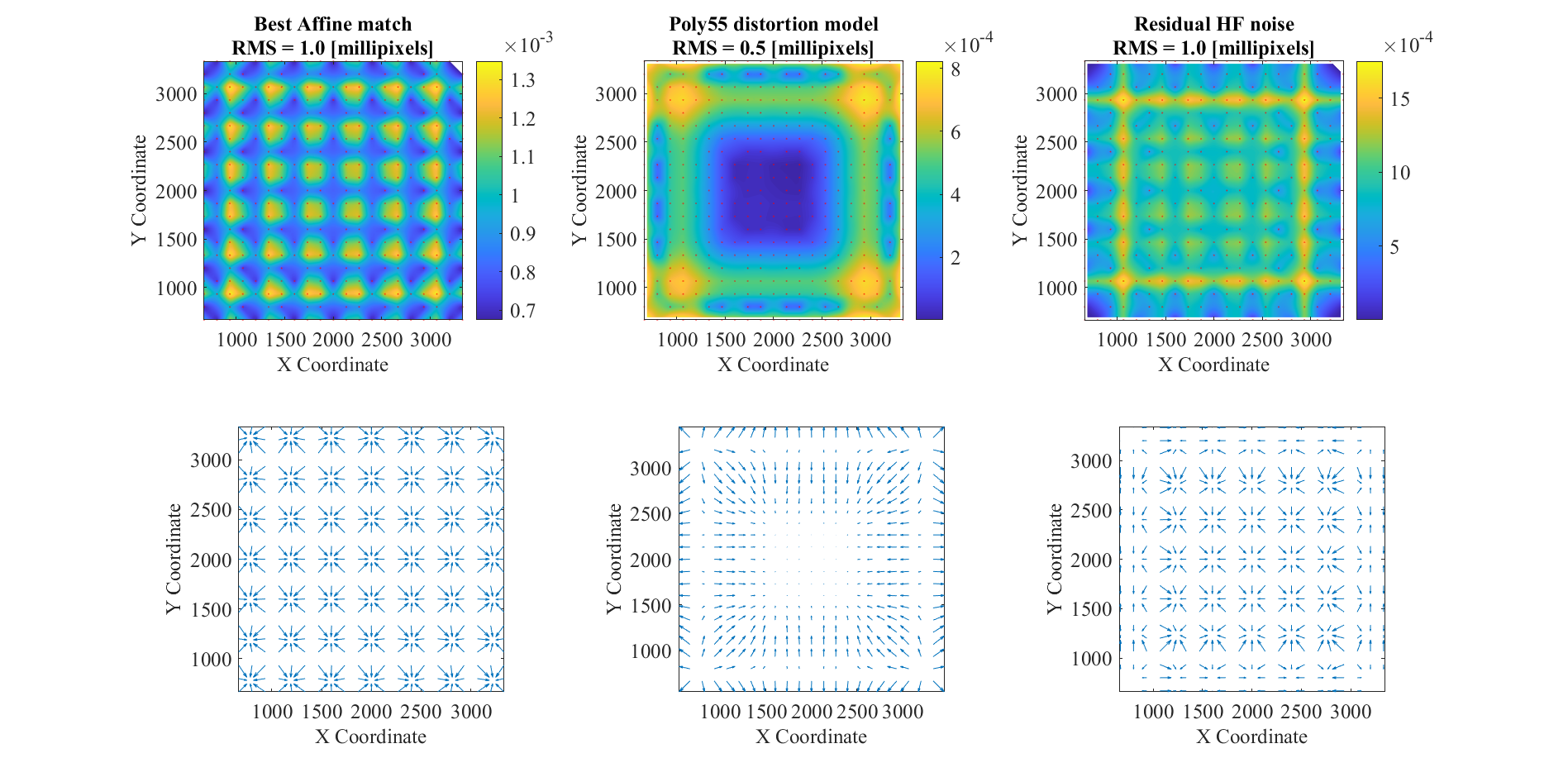}
\caption{Astrometric reduction plots for Scenario S1: 441 stars with SuperStar, no atmospheric turbulence. Left: Distortion after best affine match, Middle: Degree five distortion model, Right: Astrometric residual.}
\label{fig:appendix_S1}
\end{figure}
\begin{figure}[h!]
\centering
\includegraphics[width=1\textwidth]{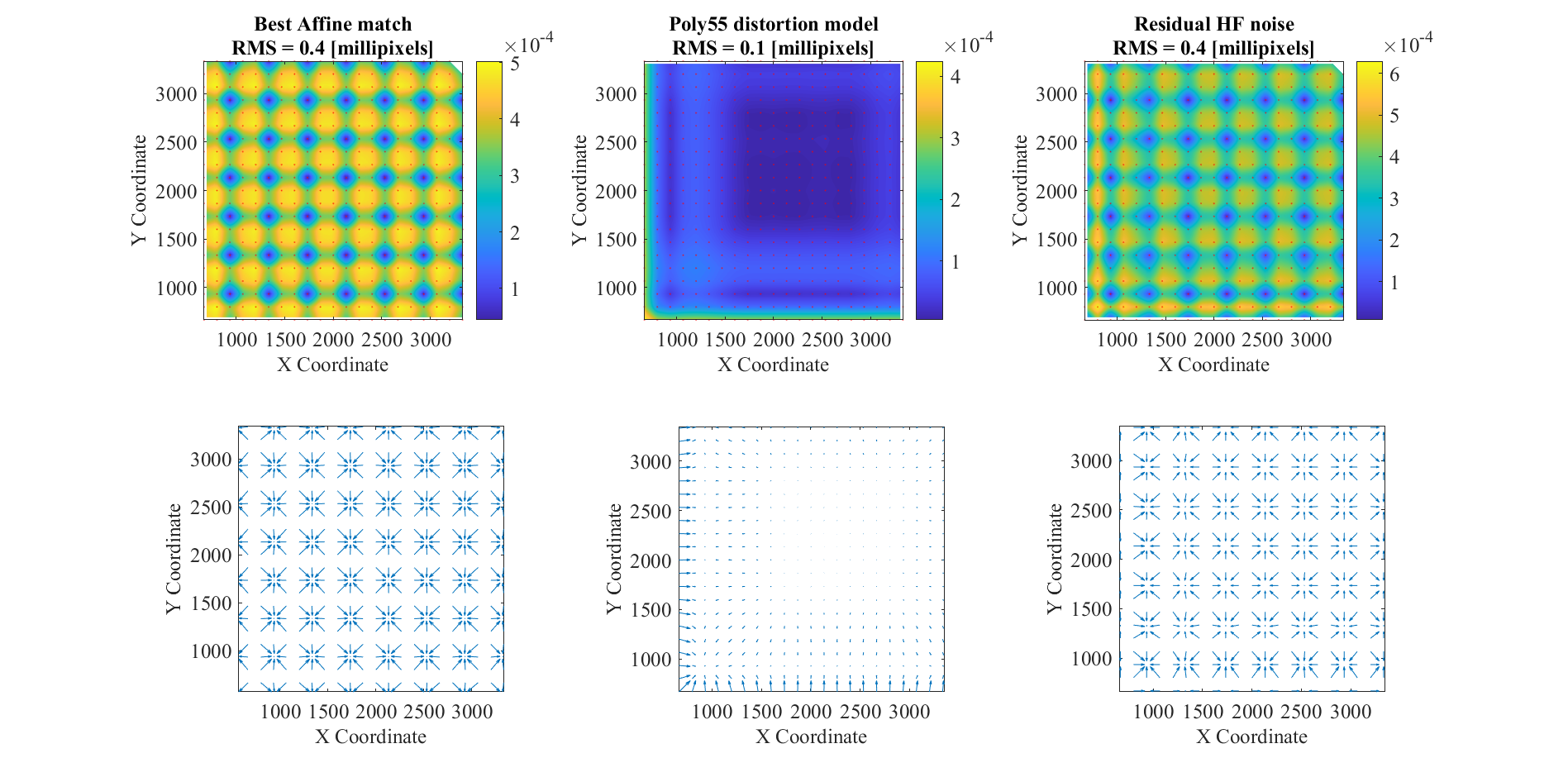}
\caption{Astrometric reduction plots for Scenario D1: 441 stars with DAOPHOT, no atmospheric turbulence. Left: Distortion after best affine match, Middle: Degree five distortion model, Right: Astrometric residual.}
\label{fig:appendix_D1}
\end{figure}
\begin{figure}[h!]
\centering
\includegraphics[width=1\textwidth]{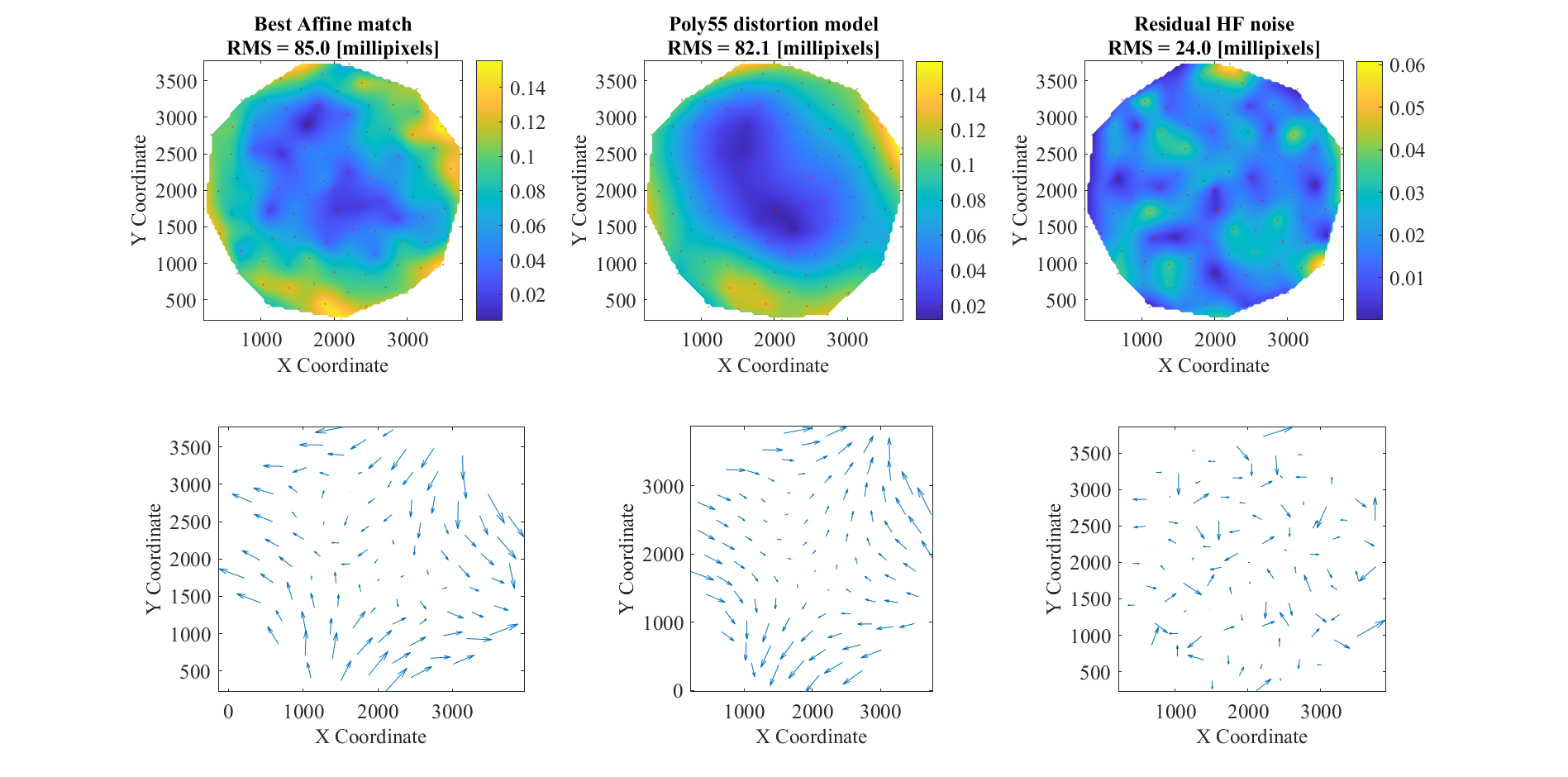}
\caption{Astrometric reduction plots for Scenario S2: 100 stars with Superstar, with atmospheric turbulence and MCAO correction. Left: Distortion after best affine match, Middle: Degree five distortion model, Right: Astrometric residual.}
\label{fig:appendix_S2}
\end{figure}

\begin{figure}[h!]
\centering
\includegraphics[width=1\textwidth]{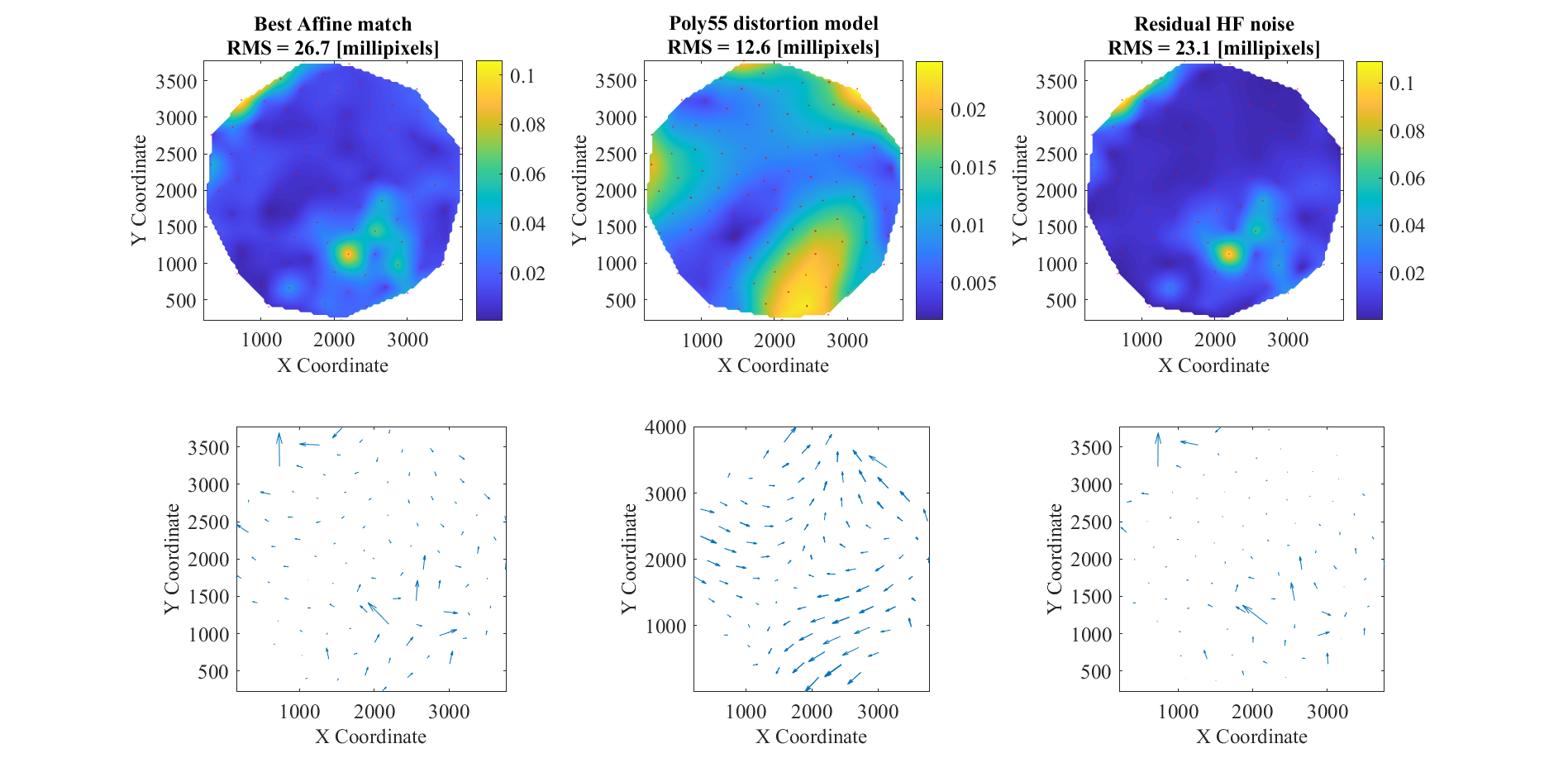}
\caption{Astrometric reduction plots for Scenario D2: 100 stars with DAOPHOT, with atmospheric turbulence MCAO correction. Left: Distortion after best affine match, Middle: Degree five distortion model, Right: Astrometric residual.}
\label{fig:appendix_D2}
\end{figure}

\begin{figure}[h!]
\centering
\includegraphics[width=1\textwidth]{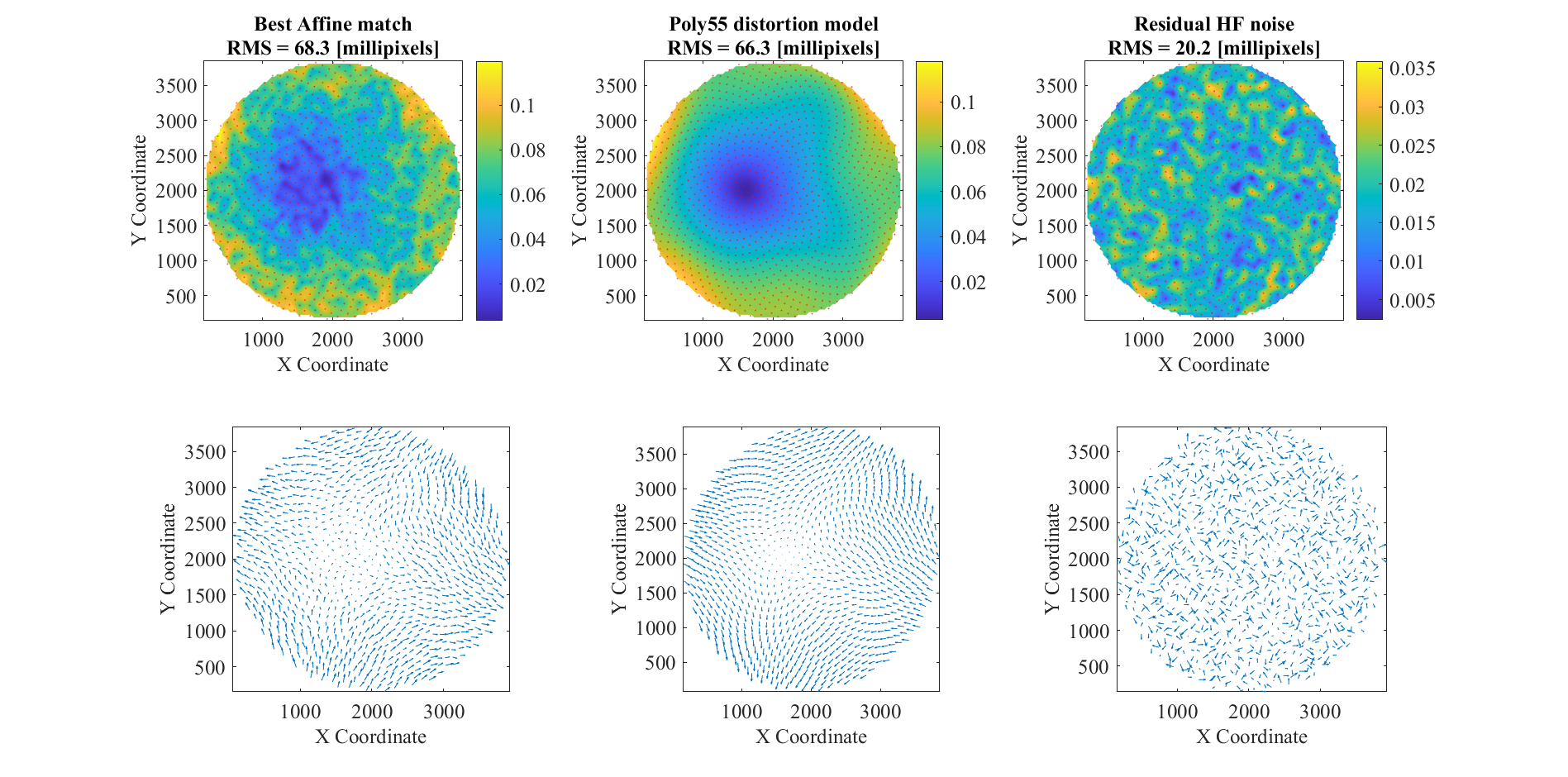}
\caption{Astrometric reduction plots for Scenario S3: 1000 stars with SuperStar, with atmospheric turbulence MCAO correction. Left: Distortion after best affine match, Middle: Degree five distortion model, Right: Astrometric residual.}
\label{fig:appendix_S3}
\end{figure}

\begin{figure}[h!]
\centering
\includegraphics[width=1\textwidth]{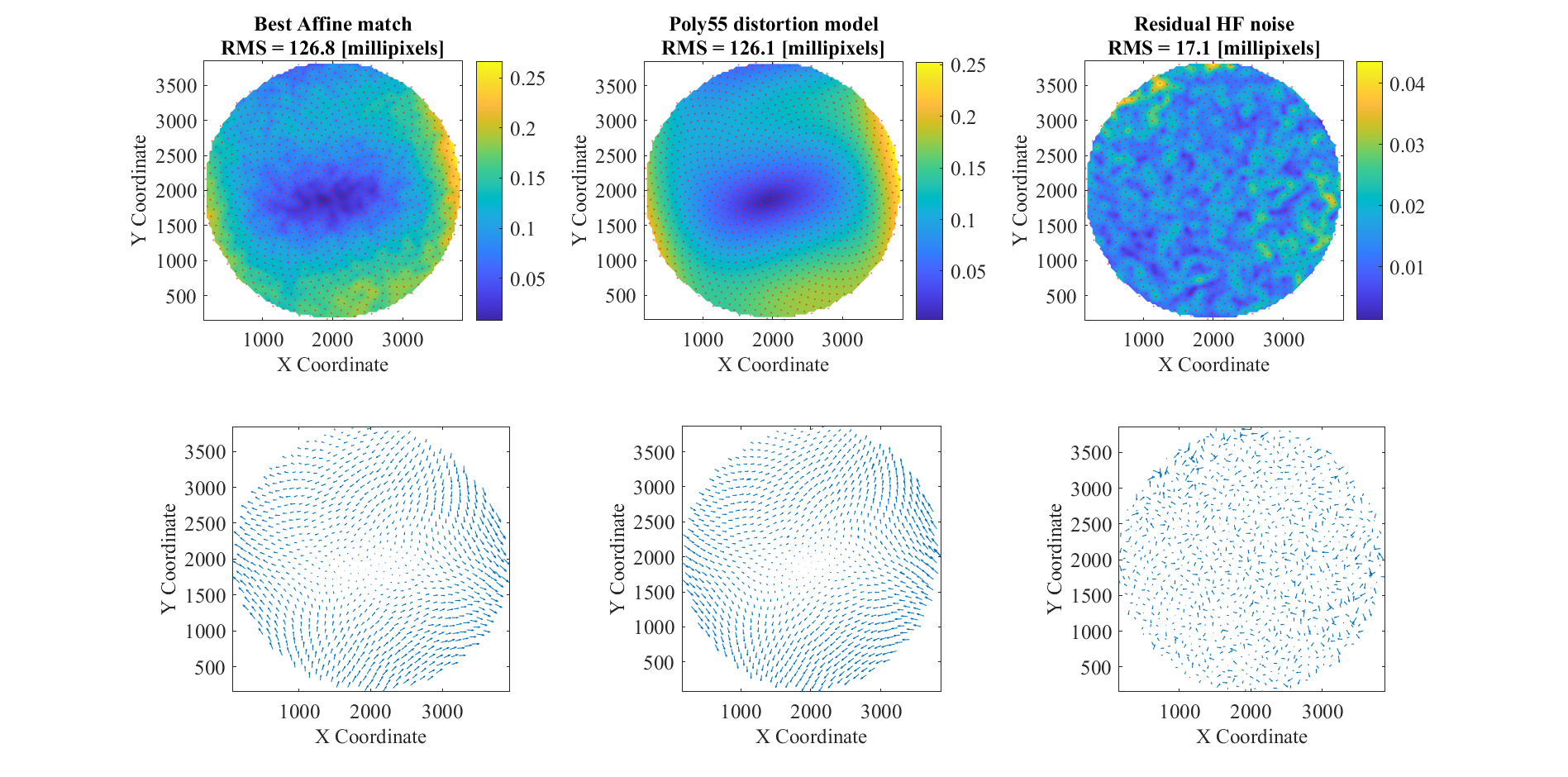}
\caption{Astrometric reduction plots for Scenario D3: 1000 stars with DAOPHOT, with atmospheric turbulence MCAO correction. Left: Distortion after best affine match, Middle: Degree five distortion model, Right: Astrometric residual.}
\label{fig:appendix_D3}
\end{figure}

\acknowledgments 
In accordance with SPIE's policy, we acknowledge the use of external AI-based tools, solely for proofreading and improving the grammar of this manuscript. MT would like to express gratitude to David Andersen from the Thirty Meter Telescope International Observatory for providing the necessary resources and support, which significantly contributed to the successful completion of this work.

\bibliography{report} 
\bibliographystyle{spiebib} 

\end{document}